\documentclass[aps, pre, onecolumn, showpacs, nobibnotes, 10pt]{revtex4-1}
\pdfoutput=1

\catcode`,\active

\catcode`\,12

\usepackage{graphicx, amsmath, amssymb, bm, mathtools}

\newcommand{\be}{\begin{equation}}
\newcommand{\ee}{\end{equation}}
\newcommand{\BE}{\begin{eqnarray}}
\newcommand{\EE}{\end{eqnarray}}
\newcommand{\D}[2]{\frac{\partial #1}{\partial #2}}
\newcommand{\DD}[2]{\frac{\partial^2 #1}{\partial #2^2}}
\newcommand{\dd}{~\mathrm{d}}
\newcommand{\ip}[1]{\langle #1 \rangle}

\usepackage{ifpdf}
\ifpdf	
	\newcommand{\ignoreThis}[1]{}
	\usepackage{color}	
\else		
	\newcommand{\ignoreThis}[1]{#1}
	\usepackage[usenames,dvips]{color}
\fi

\begin{document}

\title{The Statistics of Fixation Times for Systems with Recruitment}

\author{Tommaso Biancalani}\thanks{These authors contributed equally to this work.}
\affiliation{Department of Physics, University of Illinois at Urbana-Champaign, Loomis Laboratory of Physics, 1110 West Green Street,
Urbana, Illinois 61801-3080, USA}
\author{Louise Dyson$^*$}
\author{Alan J.~McKane}
\affiliation{Theoretical Physics Division, School of Physics and Astronomy, University of Manchester, Manchester M13 9PL, United Kingdom}

\begin{abstract} 
We investigate the statistics of the time taken for a system driven by recruitment to reach fixation. Our model describes a series of experiments where a population is confronted with two identical options, resulting in the system fixating on one of the options. For a specific population size, we show that the time distribution behaves like an inverse Gaussian with an exponential decay. Varying the population size reveals that the timescale of the decay depends on the population size and allows the critical population number, below which fixation occurs, to be estimated from experimental data.
\end{abstract}

\maketitle
\section{Introduction}
When the same phenomenon is discovered in diverse areas, it indicates that there may be an elegant and simple shared explanation. One such phenomenon is found in the following experiments in varied fields: in foraging colonies~\cite{Pasteels1987} (reviewed in~\cite{Detrain2006, *sumpter2010collective}); queueing dynamics~\cite{Becker1991}; herd investment behaviour~\cite{Scharfstein1990}; and the evolution of language~\cite{croft2000explaining} (modelling reviewed in~\cite{blythe2009generic}). All these studies have an underlying recruitment mechanism and display similar behaviours. To illustrate the general phenomenon we take, as an example, the evolution of language. 

Language is a complex and evolving system for human communication. Consider two linguistic variants, competing to become the single shared convention (\emph{i.e.} reach fixation). This competition occurs via a recruitment process, where an individual using a particular variant may induce conversational partners to also use this variant. Thus the more people using a certain variant, the more are recruited to also use it. If, by random chance, a particular variant becomes more popular, then recruitment can amplify the disparity, until fixation is reached. Other systems also display an analogous behaviour~\cite{Pasteels1987, Becker1991, Scharfstein1990, saloma2003self} and all share three basic traits: a population-based system, two (equally favourable) options, and a recruitment mechanism leading to autocatalytic amplification. In~\cite{Kirman1993} it was argued that the experiments in~\cite{Pasteels1987, Becker1991, Scharfstein1990} share a single explanation.
 
In the language of birth-death processes~\cite{Gardiner2009, *Kampen2007}, recruitment can be described using the terminology of chemical reactions: 
\be \label{IBM1}
	X+Y\xrightarrow{r=1}2X,\quad X+Y\xrightarrow{r=1} 2Y,
\ee 
where $X$ and $Y$ indicate an individual choosing one of the two options. An additional term,
\be \label{IBM2}
	X\xleftrightarrow{\epsilon} Y,
\ee 
describes the presence of spontaneous changes and its strength, $\epsilon$, is supposed small compared to that of the previous reactions. We have obtained this reaction scheme as a simplification of the Togashi-Kaneko four species model~\cite{Togashi2001} and proposed it as a description of systems with recruitment~\cite{biancalani2014noise}. An alternative method to our analytical treatment, based on the discrete time Markov chain, has been recently proposed to study these schemes in the two and three species variants~\cite{saito2014theoretical}. 

To gain intuitive understanding, one can approximate the reaction scheme by means of an expansion in the inverse of the population size, which yields the following stochastic differential equation as $\epsilon \rightarrow 0$~\cite{Biancalani2012, biancalani2014noise}:
\be\label{xeq}
	\dot x = -x + \sqrt{\frac{1-x^2}{\lambda}} \eta (t),
\ee
where $x$ denotes the difference in the concentration of individuals choosing each of the two options $X$ and $Y$, and where the dot denotes differentiation with respect to time. Here $\eta(t)$ is Gaussian white noise with zero mean and correlator 
\be
	\langle \eta(t) \eta(t') \rangle = \delta(t-t').
\ee
The parameter $\lambda>0$ is proportional to the population size and time $t$ has been rescaled by $\epsilon$~\cite{biancalani2014noise}. Under the change of variable, $x=(1-z)/2$, we recover the usual equation describing the Moran process with mutations~\cite{ewens2004mathematical}. The model also has similarities to the voter model but with the additional deterministic term, $-x$. The voter model in its most similar form can be described by $\dot x = \sqrt{1-x^2} \eta (t)$ \cite{Hammal2005}, so that $x=\pm 1$ are absorbing states of the system. Interestingly, when the system grows at a constant rate, Eq.~\eqref{xeq} can describe the voter model in rescaled time \cite{Morris2014}.

Equation~\eqref{xeq} exhibits a type of bistability in which the bistable states do not correspond to fixed points~\cite{doering1986stability}\cite{Horsthemke1984, *Popovic2013}. Other models with similar mechanisms have been discussed in the recent literature~\cite{al2005langevin, *russell2011noise, *assaf2013extrinsic, *rogers2013consensus, *remondini2013analysis, *parker2011noise}. The deterministic part of the equation has a unique stable fixed point at $x=0$, and for large values of $\lambda$ (i.e. large populations) the second term becomes negligible, and the system resides at this point. However, when $\lambda$ is smaller than some critical value ($\lambda<\lambda_c\equiv 1$), a noise-induced transition takes place~\cite{biancalani2014noise}: the noise term becomes dominant, and is maximal at $x=0$. The system is therefore driven away from the deterministic steady state at $x=0$ and towards $x=\pm 1$, where the noise term vanishes and the population consists of individuals of a single species. Once at these states, the system either displays metastability or is absorbed, according to the boundary conditions in use. If Eq.~\eqref{xeq} is being studied as a representation of the individual-based system described in \eqref{IBM1} and \eqref{IBM2} then $x$ is constrained to lie within the interval $[-1,1]$ as values outside this interval would correspond to negative population numbers. For some systems, such as in the evolution of language, a fixed population may not subsequently re-introduce a rejected linguistic variant and so absorbing boundary conditions halts the dynamical process when fixation is reached. In other systems, such as in foraging ant colonies, the process continues to be of interest after the boundary is reached, so that reflecting boundary conditions are more appropriate, and the movement between bistable states may be studied.

In this paper, we investigate the distribution of times taken for a system initialised with no bias (at $x=0$) to reach fixation (at $x=\pm 1$). For $\lambda=\lambda_c/2$, Eq.~\eqref{xeq} is exactly solvable (solution given in Eq.~\eqref{pxt}) and thus the full distribution of times may be found in terms of the derivative of a Jacobi theta function (Eq.~\eqref{ft} or Eq.~\eqref{ft2}). For general $\lambda$, we calculate an analogous, but approximate, solution (Eq.~\eqref{E:fullFT}), which is checked against simulations of the reaction scheme. Our analytical treatment indicates that the distribution of times taken to reach fixation decays exponentially at long times (Eq.~\eqref{finf} and Eq.~\eqref{E:ltFT}), with a timescale dependent on the population size. At short times, for $\lambda=\lambda_c/2$, the distribution can be approximated by an inverse Gaussian distribution (Eq.~\eqref{f0}), which captures the initial growth and its skewed maximum.

\section{Analytical treatment for the case $\lambda=\lambda_c/2$} \label{sec:solv}
\subsection{The time-dependent distribution}
We begin by finding $P(x,t)$, which is defined as the solution of the Fokker-Planck equation corresponding to Eq.~\eqref{xeq} for $\lambda=\lambda_c/2$. To proceed, we find a change of variables in Eq.~\eqref{xeq} under which the noise becomes purely additive. Since the equation is defined in the It\={o} sense, the change of variables must be performed using the It\={o} formula~\cite{Gardiner2009} which is given in this case by
\be 
	\dot{y}[x(t)] = \left[ -x(t) y'(x(t)) + \frac{1-x(t)^2}{2\lambda} y''(x(t))\right] + y'(x(t)) \sqrt{\frac{1-x(t)^2}{\lambda}} \eta(t),
\ee
so that taking $y = \arcsin(x)$, the equation becomes
\be \label{yeq}
	\dot y = \frac{1}{2}(\frac{1}{\lambda} - 2)\tan(y) + \frac{1}{\sqrt \lambda} \, \eta(t). 
\ee
Thus, for $\lambda=1/2$, the deterministic part vanishes so that Eq.~\eqref{yeq} reduces to $\dot y = \sqrt 2 \, \eta(t)$. The corresponding Fokker-Planck equation is the diffusion equation: 
\be
	\partial_t Q(y, t) = \partial^2_y Q(y,t).
\ee 
Note that since $x\in[-1,1]$, the $y$-variable is constrained to lie in the interval $[-\pi/2, \pi/2]$. 

The strategy now consists of solving the diffusion equation before reversing the transformation via $P(x,t)=Q(y,t)dy/dx$, to obtain the solution to Eq.~\eqref{xeq}. We take the system to be initially localised, $P(x,0)=\delta(x-x_0)$ and impose reflective boundary conditions at $y=\pm\pi/2$. Note that the solution of the diffusion equation cannot be simply a Gaussian because of the constraint $y\in [-\pi/2, \pi/2]$. Even though the localised system initially spreads in a Gaussian way, once it reaches $y\approx \pm \pi/2$ it accumulates at the impassable boundaries. The solution to the diffusion equation for these particular boundary conditions has been discussed in the literature both by physicists~\cite{smoluchowski1913,soskin1987} and mathematicians~\cite{karlin1960,biane2001probability}, but for convenience we reproduce it in Appendix~\ref{sec:A} (see also Ref.~\cite{polyanin_handbook_2002}). One sees that the probability density function corresponding to Eq.~\eqref{xeq} is given by
\be \label{pxt}
	P(x, t)=\frac{\theta _3\left(\arcsin(x)-\arcsin\left(x_0\right), e^{-4\,t}\right)}{\pi \sqrt{1-x^2}},
\ee
where $\theta_n (s,q)$ is the $n$-th Jacobi elliptic theta function~\cite{armitage2006elliptic}.

\begin{figure}[htpc!]
\begin{center}
    \includegraphics[width = 0.45\textwidth]{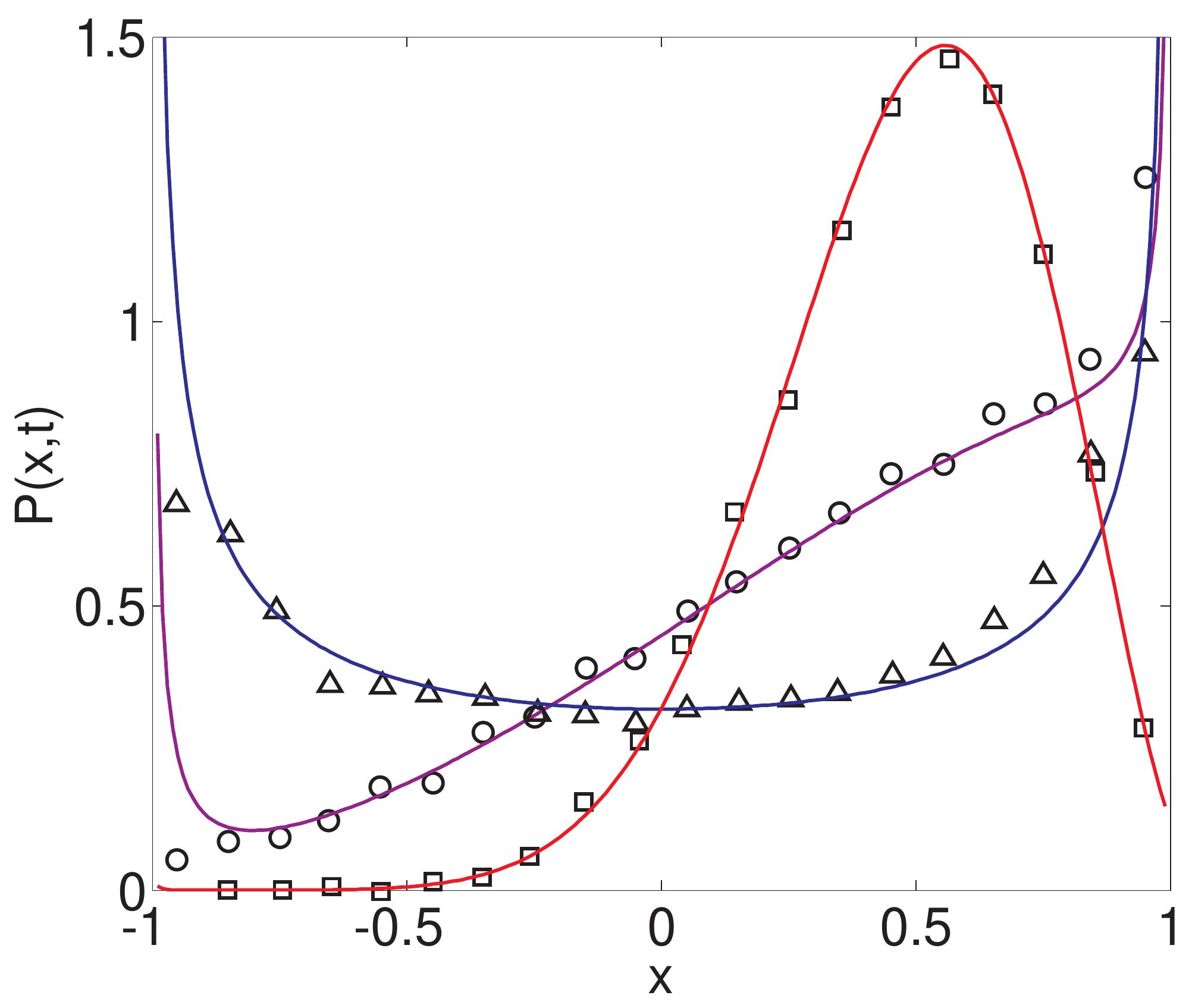} \\
\caption{The analytical expression of $P(x,t)$ (solid lines) defined in Eq.~\eqref{pxt}, is displayed against Gillespie simulations~\cite{Gillespie1977, *gillespie2013perspective} of the reaction scheme (symbols) with $N=\lambda_c/2 \epsilon$ particles and $\epsilon=10^{-2}$. The initial condition is $x_0=0.5$ and we show results for three different times: $t=0.05$ (red line, squares), $t=0.2$ (purple line, circles), $t=2$ (blue line, triangles). \label{fig:pxt}}
\end{center}
\end{figure}

The distribution~\eqref{pxt} is positive and normalised in the interval $[-1,1]$. It diverges at $x=\pm 1$ but this is not a sign of singular behaviour as the probability $P(x,t)\Delta x$, evaluated on $x=\pm 1$, tends to zero as $\Delta x \to 0$, meaning that the system is well behaved at these points. As $t\to\infty$, $P(x,t)$ relaxes to the stationary distribution 
\be
	P_s(x)=\pi^{-1} \left(1-x^2\right)^{-1/2},
\ee 
which indicates that the system spends most of its time in proximity to $x\approx\pm 1$. The dynamics from the localised initial condition to the final stationary distribution is shown in Fig.~\ref{fig:pxt}.

\subsection{Time statistics}
The change of variables, $y = \arcsin(x)$, is also instrumental in finding an exact solution for the distribution $f(T)$ (still in the case where $\lambda=\lambda_c/2$) of the time $T$ taken for the system, initialised at $x=0$, to reach one of the two states, $x = \pm 1$. This is equivalent to initialising the system in the $y$ variable at $y=0$, and imposing absorbing boundary conditions at $y=\pm\pi/2$. These latter conditions ensure that the dynamics cease when the system reaches one of the states $x=\pm 1$ for the first time. Performing these calculations in an analogous way to before (see Appendix~\ref{sec:B}) gives
\be \label{ft}
	f(T) = \frac{2}{\pi} \,\theta'_1(0, e^{-4 T}),
\ee
where we used the notation $\theta_1'(s,q) \equiv \partial_s\theta_1(s,q)$.
\begin{figure}[htpc!]
\begin{center}
	\includegraphics[width=0.45\textwidth]{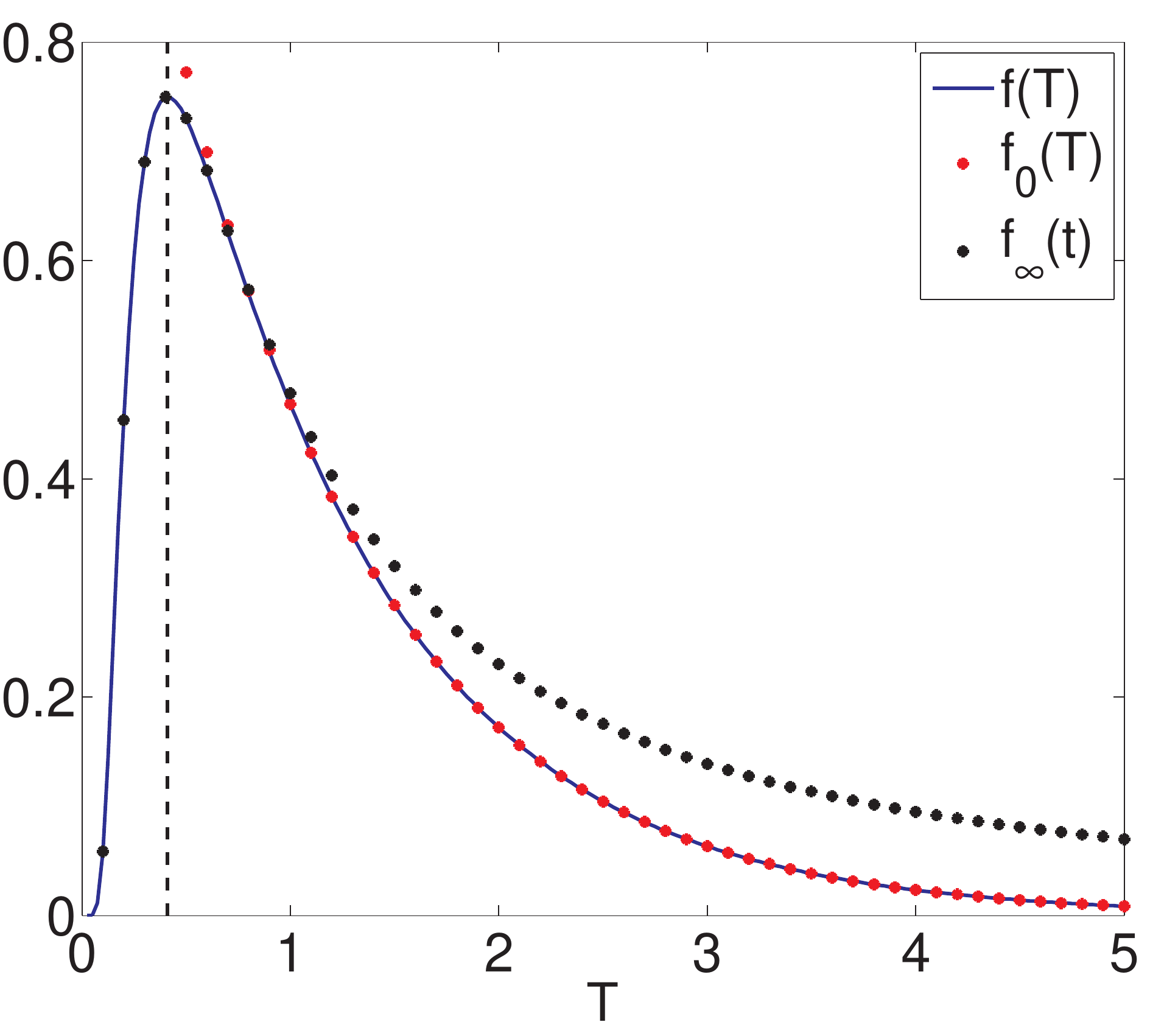} \\
\caption{The function $f(T)$ (solid line), Eq.~\eqref{ft}, is shown against its asymptotic approximations: $f_0(T)$ (black dots), Eq.~\eqref{f0}, and $f_\infty(T)$ (red dots), Eq.~\eqref{finf}. The vertical line indicates the maximum point, $T_m = \pi^2/24$, of $f_0(T)$.\label{fig:ft}}
\end{center}
\end{figure}

The distribution~\eqref{ft} (Fig.~\ref{fig:ft}) is unimodal, so that the dominant timescale is given by the single maximum of the distribution, and skewed, suggesting that the maximum can significantly differ from the mean. We may further understand this equation by carrying out an asymptotic analysis. We use an expansion for the derivative of the Jacobi elliptic theta function, $\theta_1'(s,q)$, which holds for $s=0$ and small $q$. At first order this is~\cite{armitage2006elliptic} 
\be \label{exp}
	\theta'_1(0,q)\approx 2 q^{\frac{1}{4}}.
\ee 
Applying this to Eq.~\eqref{ft} leads to an approximation for large times
\be \label{finf}
	f(T) \approx f_\infty(T) = \frac{4}{\pi} e^{-T}, 
\ee
which indicates that the decay is exponential.

Asymptotics for short times can also be obtained, but we need to first apply the Jacobi imaginary transform~\cite{armitage2006elliptic} so that we may again use the previous expansion for small $q$. The transform is 
\be	
	\theta_1(y, e^{i \pi \alpha}) = \sqrt{\frac{-i}{\alpha}}\,\exp{\left(\frac{y^2}{\pi i \alpha}\right)}\,\theta_1(-\alpha^{-1} y, e^{- \frac{i \pi}{\alpha}}).
\ee
Setting $\alpha = 4 i T/\pi$, taking the derivative with respect to $y$, and evaluating it at $y=0$, yields an alternative expression for Eq.~\eqref{ft}:
\be \label{ft2}
	f(T ) = \frac{\sqrt \pi}{4} T^{-\frac{3}{2}}\, \theta'_1(0, e^{-\frac{\pi ^2}{4T}}).
\ee
The advantage of this form is that $T$ now appears in the denominator of the exponential allowing us to again apply the expansion, Eq.~\eqref{exp}, to Eq.~\eqref{ft2} to give, for short times,
\BE
	f(T )\approx f_0(T) &=& \sqrt{\frac{\pi}{4 T^{3}}}\,\exp{\left( {-\frac{\pi ^2}{16 T}} \right) } \nonumber \\
&\propto& \frac{\bar z}{\sqrt{D T^3}}\,\exp{- \left( \frac{\bar z^2}{4 D T} \right) }.
\label{f0}
\EE

This last expression, when normalised, is already known in a different context. It is the inverse Gaussian distribution and gives the statistics of times taken by a one-dimensional Brownian particle (here with diffusion coefficient $D=1$) starting from the origin in a semi-infinite system, $[\bar z, \infty)$, to reach an absorbing boundary at $\bar z$ \cite{redner2001guide} (here $\bar z = \pi/2$). The factor $T^{-3/2}$ is a consequence of the one-dimensional nature of the Brownian motion \cite{redner2001guide} and gives the leading order behaviour for absorbing states $\bar z$ far from the origin. The presence of the exponential reveals that the absorption times are of order $T \sim O(\bar z^2/D)$ and a computation of the position of the maximum value gives 
\be
	T_m = \frac{\bar z^2}{6D} \approx 0.41. 
\ee
The asymptotic approximations~\eqref{finf} and~\eqref{f0} and the position of the maximum, $T_m$, are shown in Fig. \ref{fig:ft}.

\section{Analytical treatment for a general $\lambda$}\label{sec:gen}
Returning to Eq.~\eqref{xeq} with general $\lambda$, we wish to investigate the time taken for an unbiased system to pick one of the two states $x=\pm 1$. We cannot approach this in a similar way to the earlier $\lambda = \lambda_c/2$ case as the deterministic part of Eq.~\eqref{yeq} does not vanish in this regime. However, the distribution of times can still be obtained by separation of variables~\cite{Horsthemke1984}. We begin by writing down the backwards Fokker-Planck equation~\cite{Kampen2007} corresponding to Eq.~\eqref{xeq}:
\begin{align} \label{bfp}
	\D{G}{T} = -x \D{G}{x} +  \frac{\left(1-x^2\right)}{2\lambda} \DD{G}{x},
\end{align}
with absorbing boundary conditions, $G(x=1,T) = 0,$ and $G(x=-1,T) = 0$.
The function $G(x,T)$ denotes the density of probability that the system has not escaped from the interval $[-1,1]$ after a time $T$. The required  distribution $f(T)$ is thus given by the rate $-\partial G/\partial T$ that the system leaves the domain~\cite{Gardiner2009,Kampen2007}. The initial condition is given by $G(x,0) = 1$, since a system initialised in the domain must by definition remain in the domain at time $T=0$. 

Equation~\eqref{bfp} may be solved giving rise to series solutions in terms of the eigenfunctions of the right-hand side. We describe the method here, and refer the reader to the attached CDF / \textit{Mathematica} file for the explicit expressions of some of the resulting quantities. Searching for separable solutions and using the boundary conditions reveals a discrete set of eigenvalues, $F_n$, which give the inverse of the timescales at which the profiles given by the corresponding (unnormalised) eigenfunctions $v_n$ decay. The eigenvalues and eigenfunctions are given by
\begin{equation}
	\begin{split}
	F_n &= \frac{n(1+n-2\lambda)}{2\lambda}, \\ 	
	v_n &= \sum_{k=0}^\infty L_{k,n,\lambda} \sin((n+2k)\phi),
	\end{split}
\end{equation}
for $n\in \mathbb{N}$ and $\phi = \arccos(x)$. The expression for the coefficients $L_{k,n,\lambda}$ is given in the attached CDF / \textit{Mathematica} file.

The solutions are therefore given by 
\begin{align} \label{ansatz}
	G(\phi,T) = \sum_{n=0}^\infty C_n e^{-F_n T} v_n(\phi),
\end{align}
where the $C_n$ are constants that are determined using the initial condition $G(\phi,0) = 1$. The initial condition is applied by using the orthogonality of the eigenfunctions~\cite{Horsthemke1984}, $v_n(\phi)$, of the backward Fokker-Planck equation, under the inner product defined by 
\begin{align} \label{ip}
	\ip{v_n,v_m} = \int_0^\pi \sin\phi P_s(\cos\phi)v_n(\phi)v_m(\phi) \dd \phi,
\end{align}
where 
\be
	P_s(\cos\phi) = \mathcal N (\sin\phi)^{2(\lambda-1)}
\ee 
is the stationary distribution of Eq.~\eqref{xeq}~\cite{biancalani2014noise},
and $\mathcal N$ is a normalisation constant. Since the eigenfunctions are orthogonal with respect to the inner product~\eqref{ip}, $\ip{v_n,v_m}=0$ for $n\ne m$. Thus taking the inner product of the initial condition (Eq.~\eqref{ansatz} at $T=0$) with $v_n$ reveals that 
\be
	\ip{G(\phi,0),v_n} = C_n \ip{v_n,v_n},
\ee
and gives an expression for $C_n$. 

\begin{figure}[htpc!]
\begin{center}
	\includegraphics[width=0.5\textwidth]{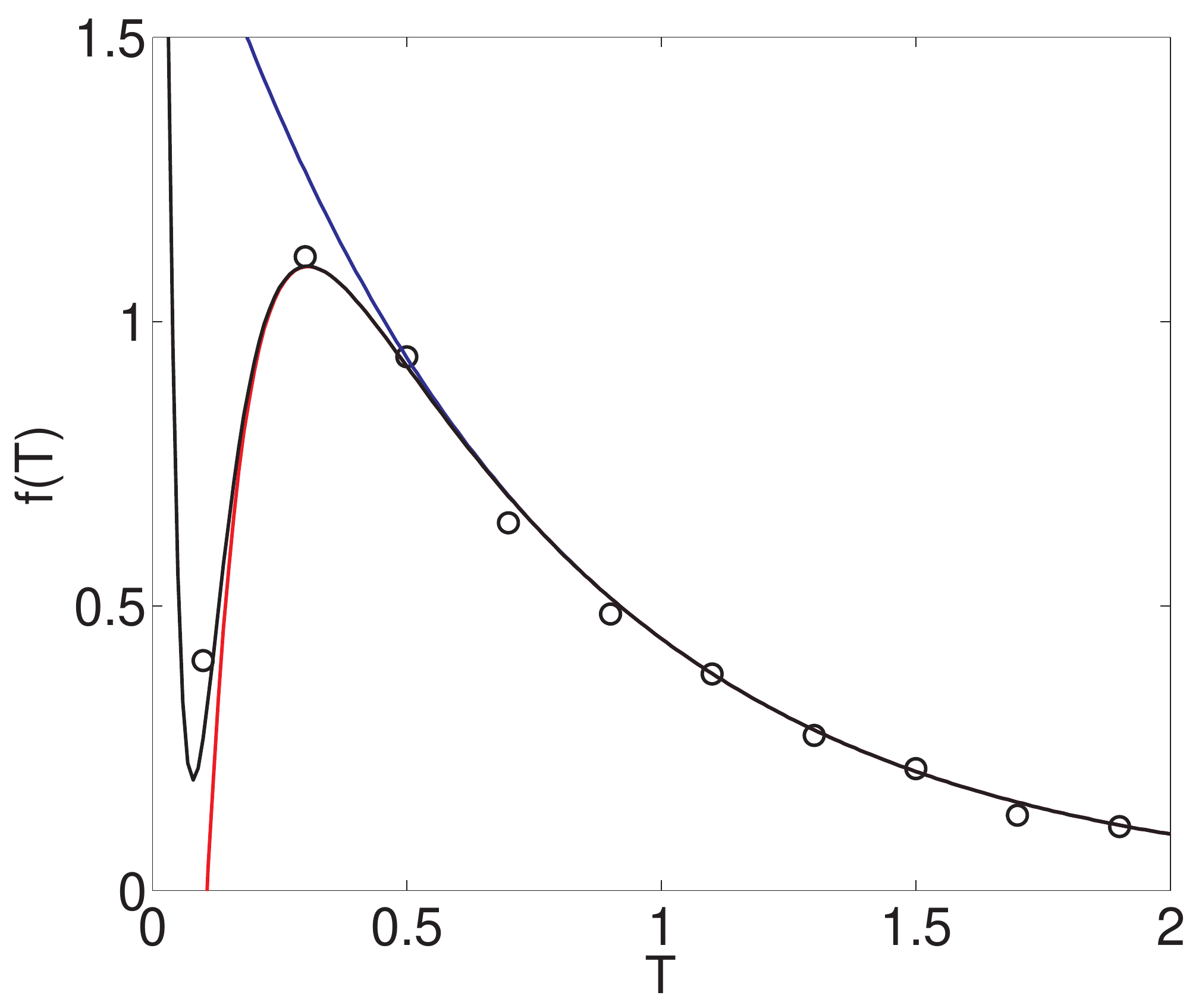} \\
\caption{\label{fig:change_terms_lam_0_6} Equation \eqref{E:fullFT} (solid lines) is compared to simulations as described in Fig.~\ref{fig:pxt} (symbols) for one (blue line), two (red line) and three (black line) terms in the summation. Simulations are averaged over $2500$ runs with $\epsilon = 10^{-3}$ and $\lambda = 0.6$, initialised at $x = 0$ and run until $x = \pm 1$.}
\end{center}
\end{figure}

Thus the distribution of fixation times is given by 
\begin{align}
f(T) = \sum_n \frac{\ip{1,v_n}}{\ip{v_n,v_n}} F_n \sum_{k=0}^\infty (-1)^{k+\frac{n-1}{2}} L_{k,n,\lambda} e^{-F_n T}. \label{E:fullFT}
\end{align}
Whilst $\ip{1,v_n}$ may be found exactly, and is zero for even values of $n$, the normalisation terms, $\ip{v_n,v_n}$, are given by an infinite summation. However, since the summand is small for all except the first term, we may take only the very first term in the sum and neglect the others. Expressions for these quantities are given in the attached CDF / \textit{Mathematica} file. The infinite $n$ summation in Eq.~\eqref{E:fullFT} is also not exactly computable and we therefore truncate the summation and observe that increasing the number of terms taken in the sum leads to a better approximation for small values of $T$ (Fig.~\ref{fig:change_terms_lam_0_6}).

\begin{figure}[htpc!]
\begin{center}
	\includegraphics[width=0.5\textwidth]{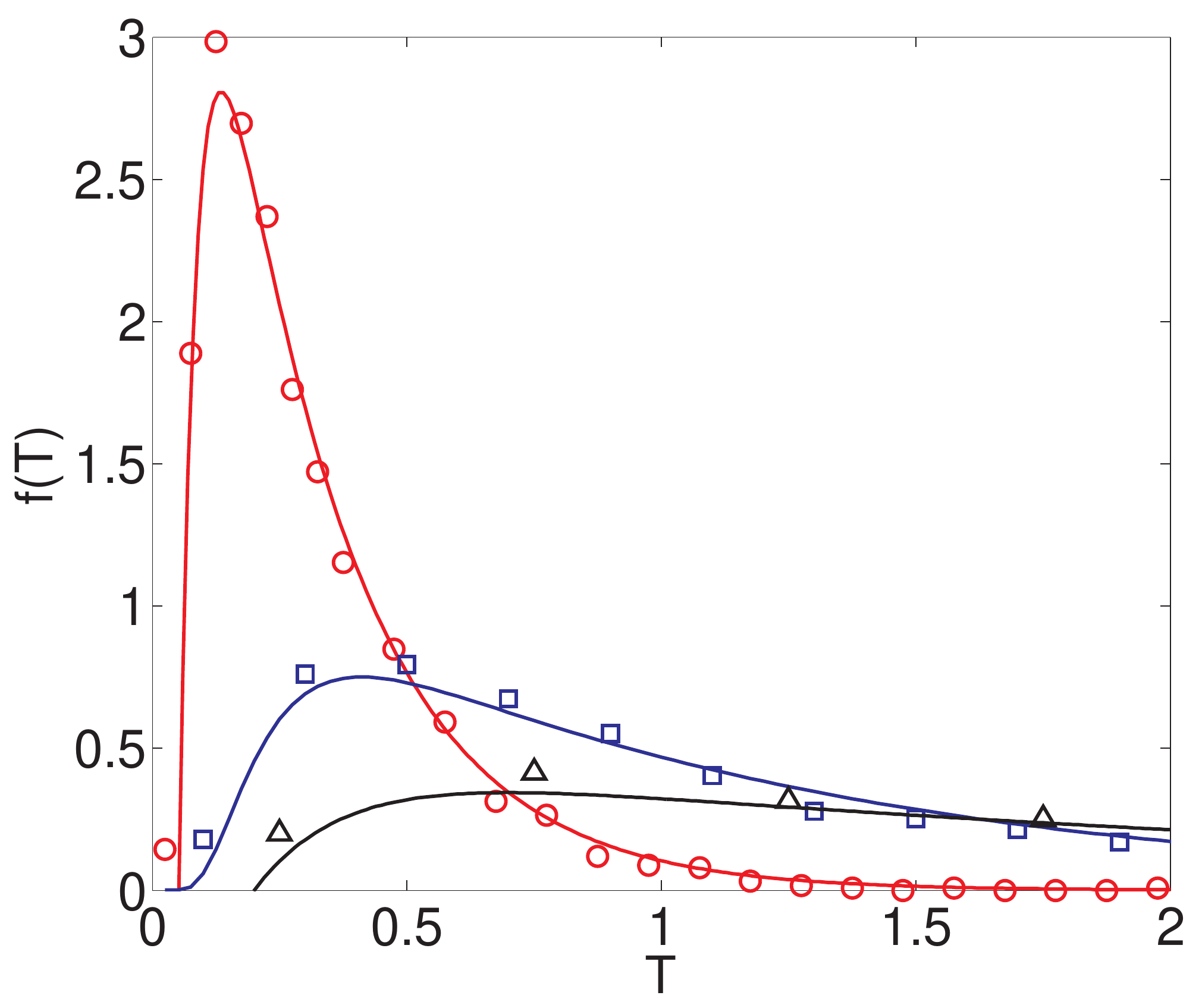} \\
\caption{\label{fig:change_lam_eps_1000} The first two terms of Eq.~\eqref{E:fullFT} (solid lines) is compared to simulations~\cite{Gillespie1977, *gillespie2013perspective} as described in Fig.~\ref{fig:pxt} (symbols) for  $\lambda = 0.2$ (red), $\lambda = 0.5$ (blue) and $\lambda = 0.7$ (black). Simulations are averaged over $2500$ runs with $\epsilon = 10^{-3}$, initialised at $x = 0$ and stopped when $x = \pm 1$.}
\end{center}
\end{figure}

Taking the first two terms in the sum gives a good fit to simulations for a range of $\lambda$ values (Fig.~\ref{fig:change_lam_eps_1000}). A better fit is given for smaller values of $\lambda$, \emph{i.e.} where there are fewer individuals in the population, since our approximation of taking $\epsilon \rightarrow 0$ to give Eq.~\eqref{xeq} is more accurate at smaller population sizes. The distribution found in the general $\lambda$ case is qualitatively similar to that in the exactly solvable case, and is skewed for all values of $\lambda$,  similarly to the $\lambda=\lambda_c/2$ case. 

Taking only the leading order term in the summation, we obtain
\begin{align}\label{E:ltFT}
	f(T) \approx -\frac{2 \Gamma \left(\frac{5}{2}-\lambda \right)}{\sqrt{\pi } \lambda^2 \Gamma (-\lambda)} e^{\frac{\lambda-1}{\lambda}T},
\end{align}
which fits the tail of the distribution (Fig.~\ref{fig:change_terms_lam_0_6}) and recapitulates the long-time approximation found in Eq.~\eqref{ft}. It is therefore clear that the distribution has an exponentially decaying tail, with a faster decay for smaller $\lambda$.

\section{Conclusion}
In this paper, we have shown that for $\lambda=\lambda_c/2$, Eq.~\eqref{xeq} can be solved and the statistics of fixation times can be found as a differentiated Jacobi elliptic theta function. An asymptotic analysis reveals that, for short times, this distribution behaves as an inverse Gaussian distribution but with an exponential decay at long times. Our analysis for a general $\lambda$ shows that the timescale of the decay is $\lambda/(\lambda-1)$ (Eq.~\eqref{E:ltFT}), which provides an experimental prediction for estimating the rescaled population size, $\lambda$, thus quantifying the distance of the system from the critical size, $\lambda_c$.

Since the model we have studied is very simple, it is worth asking if these predictions are applicable to experimental results~\cite{Pasteels1987, Becker1991, Scharfstein1990, saloma2003self, helbing2003lattice}. Equation~\eqref{xeq} displays some general features that we expect to be ubiquitous: the skewness of the distribution, the presence of a single maximum and the exponential tail. For short times, we have shown that for a specific choice of $\lambda$ the times are distributed according to an inverse Gaussian distribution. Do we expect this to hold for general $\lambda$? Our series solution provides a more general expression to fit the data. Furthermore, agreement between theory and experiment might be obtained by using generalisations of the inverse Gaussian distribution, such as the Levy or Gamma distributions.

Future research directions include exploring new fields of applicability such as the chiral symmetry-breaking observed in chemical and biological molecules~\cite{saito2013colloquium}. This has an autocatalytic mechanism and is relevant in the field of astrobiology.

\vspace{10pt}
\begin{acknowledgments}
T.B. ackowledges Nigel Goldenfeld for useful discussions concerning homochirality. L.D. was supported under EPSRC grant EP/H02171X. T.B. acknowledges partial support from the National Aeronautics and Space Administration through the NASA Astrobiology Institute under Cooperative Agreement No. NNA13AA91A issued through the Science Mission Directorate.
\end{acknowledgments}

\appendix
\section{Solution of the diffusion equation}\label{sec:A}
We begin by computing the time-dependent distribution, $P(x,t)$, of the equation:
\be \label{sm_xeq}
	\dot x = -x + \sqrt 2 \sqrt{1-x^2}\, \eta(t),
\ee
where $\eta(t)$ is Gaussian noise with zero mean and correlator,
\be
	\langle\eta(t)\eta(t')\rangle=\delta(t-t'),
\ee
and Eq.~\eqref{sm_xeq} follows from the main text by assigning $\lambda=1/ 2$. The corresponding Fokker-Planck equation is
\be \label{sm_FP}
	\D{P}{t} = \D{}{x}\left(x P\right) + \DD{}{x}\left((1-x^2)P\right).
\ee

We imagine that we start with the system localised at a certain $x_0$, and look for a solution that is positive and normalised in the $x$ domain~\cite{Gardiner2009}: 
\be \label{icbc1}
	\begin{split}
	\int_{-1}^1 dx\, P(x,t) = 1, &\quad P(x,t) > 0,\\ 
	P(x,0) &= \delta(x-x_0).
	\end{split}
\ee
Under the change of variable $y=\arcsin(x)$, Eq.~\eqref{sm_FP} becomes the diffusion equation. Translated into the $y$-variable these conditions read:
\be
	\int_{-\frac{\pi}{2}}^\frac{\pi}{2} dy\,Q(y,t) = 1, \quad Q(y,t) > 0,
\ee
and
\be
	\begin{split}
	Q(y,0) & = P(x,0) \frac{dx}{dy} =\\
	&= \delta(\sin(y) - \sin(y_0)) \cos(y) = \delta(y-y_0),
	\end{split}
\ee
where $y_0 = \arcsin(x_0)$ and the last equality holds because $y$ is restricted to $[-\pi/2, \pi/2]$. 

To solve the diffusion equation, we go over to a Fourier representation. We start by recalling that any function $f(y)$, defined in an interval of length $l$, admits a representation as a Fourier series~\cite{polyanin_handbook_2002}: 
\be
	f(y)=\sum _{k=-\infty }^{+\infty } \exp\left( i k\frac{2\pi}{l} y\right) f_k.
\ee
Since in our case the domain is an interval of length $\pi$, the probability density function $Q(y,t)$ can be rewritten as:
\be
	Q(y, t)=\sum _{k=-\infty }^{+\infty } e^{i 2 k y} a_k(t).
\ee
Inserting this into the diffusion equation gives a linear equation for the Fourier coefficients. Once solved, we have: 
\be
	a_k(t)=a_k(0) e^{-4k^2 t}.
\ee 
The coefficients at the initial time, $a_k(0)$, are determined using the initial condition and the identity:
\be
	\delta \left(y-y_0\right)=\frac{1}{2\pi }\sum _{k=-\infty }^{+\infty } \exp(i k \left(y-y_0\right)).
\ee 
This shows that $a_k(0)=\pi^{-1}e^{-2 i k y_0}$. The series for $Q(y, t)$ can now be summed and becomes:
\be
	\begin{split}
	Q(y, t)=&\pi^{-1}\sum _{k=-\infty }^{+\infty } e^{i k 2 \left(y-y_0\right)} e^{-4k^2 t} \\
	=& \pi^{-1} \theta _3\left(y-y_0,e^{-4 t}\right),
	\end{split}
\ee 
where the function $\theta_n(x,q)$ is the $n$-th elliptic theta function, with the conventions adopted in~\cite{armitage2006elliptic}. Changing back to the $x$ variable, by $P(x,t) = Q(y,t) dy/dx$, yields the time-dependent probability density function:
\be \label{sm_pxt}
	P(x, t)=\frac{\theta _3\left(\arcsin(x)-\arcsin\left(x_0\right), e^{-4\,t}\right)}{\pi \sqrt{1-x^2}}. 
\ee

\section{The time statistics for $\lambda=\lambda_c/2$}\label{sec:B}
We need to solve the diffusion equation with absorbing boundary conditions at $y=\pi/2$ and $y=-\pi/2$. The system is initialised at $y_0=0$. We can carry out an analogous calculation to the one of Appendix~\ref{sec:A}. More simply, the solution of the diffusion equation for these boundary conditions can be found in the literature~\cite{polyanin_handbook_2002}. In either case:
\be
	\begin{split}
	Q(y,t) =& \frac{2}{\pi} \sum_{n=1}^{\infty} \sin\left[ n\left(y + \frac{\pi}{2}\right)\right] \sin\left( n\frac{\pi}{2}\right) e^{-n^2 t} =\\ 
	=& \frac{2}{\pi} \sum_{n=1}^{\infty}(-1)^n \sin\left[ \left(2n+1\right)\left(y + \frac{\pi}{2}\right)\right]  e^{- 4\left(n+\frac{1}{2}\right)^2 t},
	\end{split}
\label{Q_y_t}
\ee
and following \cite{armitage2006elliptic}, we immediately recognise that:
\be
	Q(y,t) = \frac{1}{\pi} \theta_1\left( y+\frac{\pi}{2}, e^{-4t}\right).
\label{theta_one}
\ee
The integral of $Q(y, t)$ over the $y$-domain represents the probability that $y$ remains in $[-\pi/2, \pi/2)$ after a time $t$. This is the probability that $y$ has not yet reached the absorbing boundary after a time $t$. Rephrased again, it is the probability that the time at which $y$ is absorbed, $T$, is greater than $t$. We denote this probability by:
\be
	\text{Prob}(T>t) = \int_{-\pi/2}^{\pi/2}dy\,Q(y,t).
\ee
Since $\text{Prob}(T > t)$ is the cumulative distribution function of $T$, the statistics of jumps is readily
obtained by differentiation:
\be 
	\begin{split}
	f(T) =& - \partial_t \,\text{Prob}(T>t)\bigg|_{t=T} = - \int_{-\pi/2}^{\pi/2}dy\,\partial_t Q(y,t)\bigg|_{t=T} \\
	=& \partial_y Q(y, T) \bigg|_{\pi/2}^{-\pi/2},
	\end{split}
\ee
where to simplify the last integral we have used the fact that $\partial_t Q = \partial_y^2 Q$. Noting from Eqs.~\eqref{Q_y_t} and \eqref{theta_one} that 
$\theta'_1(\pi, e^{-4 t}) = - \theta'_1(0, e^{-4 t})$, we see that
\be
f(T) = \frac{2}{\pi} \,\theta'_1(0, e^{-4 T}).
\ee

\bibliographystyle{apsrev4-1}

\end{document}